\documentclass[12pt]{article}
\usepackage{amsmath,amssymb}
\renewcommand{\a}{\alpha}

\newcommand{\bs}{\bigskip}
\newcommand{\dt}{\cdot}

\newcommand{\e}{\varepsilon}
\renewcommand{\i}{\infty}
\renewcommand{\k}{\kappa}
\renewcommand{\l}{\lambda}

\newcommand{\mb}{\mbox}
\renewcommand{\ni}{\noindent}
\newcommand{\p}{\partial}

\begin{document}
\setlength{\baselineskip}{16pt}

\begin{center}
{\large\bf Self-Similar Intermediate Asymptotics for a}\\
{\large\bf Degenerate Parabolic Filtration-Absorption
Equation}

\vspace{.5 truein}
{\sc g. i. barenblatt,${}^*$ m. bertsch,${}^\S$ a. e.
chertock,${}^*$} and {\sc v. m. prostokishin${}^\dagger$}

\bs

${}^*$Department of Mathematics and\\
Lawrence Berkeley National Laboratory\\
University of California\\
Berkeley, California 94720 USA;

\bs

${}^\S$Department of Mathematics\\
University of Rome `Tor Vergata'\\
Via della Ricerca Scientifica\\00133 Rome, Italy; and

\bs

${}^\dagger$P. P. Shirshov Institute of Oceanology\\
Russian Academy of Sciences\\
36 Nakhimov Prospect\\
117218 Moscow, Russia
\end{center}

\vspace{.75 truein}
\begin{quote}
{\footnotesize\bf The equation}
$$
\p_tu=u\p^2_{xx}u-(c-1)(\p_xu)^2
$$
{\footnotesize\bf is known in literature as a qualitative
mathematical model of some biological phenomena. Here this
equation is derived as a model of the groundwater flow in a water
absorbing fissurized porous rock, therefore we refer to
this equation as a  filtration-absorption
equation. A family of self-similar solutions to
this equation is constructed. Numerical investigation of
the evolution of non-self-similar solutions to the Cauchy
problems having compactly supported initial conditions is
performed. Numerical experiments indicate that the
self-similar solutions obtained represent intermediate
asymptotics of a wider class of solutions when the
influence of details of the initial conditions disappears
but the solution is still far from the ultimate state:
identical zero. An open problem caused by the nonuniqueness
of the solution of the Cauchy problem  is discussed.}
\end{quote}

\section{A derivation of the filtration-absorption
equation}

A permeable rock layer on a horizontal impermeable bed is
considered. It is well known that the gently sloping
groundwater flows in a purely porous medium with
infiltration are described by the Boussinesq equation
(see refs.~(1--3)):
$$
m\p_th=\k \big( h\p^2_{xx} h+(\p_x h)^2\big)+q \ .
\eqno[1]
$$
Here $h$ is the groundwater level, $t$ is the time, $x$
is the horizontal space coordinate along the impermeable
bed, $q$ is the intensity of the
groundwater inflow or outflow. Furthermore, $\k=\rho
gk/\mu$ is a joint property of the pair-rock/fluid,
assumed here to be constant, $k$ is the rock
permeability, $m$ the rock porosity, $g$ the gravity
acceleration, $\rho$ and $\mu$ are fluid density and
dynamic viscosity.

Assume now that when the fluid level is decreasing, some part of
the fluid remains being absorbed by rock, e.g.~due to capillary
imbibition to the micropores. Then for a fixed fluid particle the
rate of fluid absorption $q$ will be proportional to the {\it
individual} time derivative of the fluid level $dh/dt$, so that
$$
q=\a m \ \frac{dh}{dt} =\a m(\p_t h+u_p\p_x h ) \ .
\eqno[2]
$$
Here $\a$ is also a joint rock/fluid property which we also assume
here to be constant, and $u_p$ is the actual fluid velocity.

According to the Darcy law the filtration velocity (fluid flux per
unit area) is equal to
$$
u_f=- \ \frac{\rho g k}{\mu} \ \p_xh \ . \eqno[3]
$$
The actual fluid velocity is different (see (1))
from the filtration velocity $u_f$. For a purely porous medium
$u_p=u_f/m$. It is very important that if the rock is fissurized,
i.e.~contains a connected network of cracks, then
$$
u_p=\frac{u_f}{m_1} \eqno[4]
$$
where $m_1$ ({\it ``fissure porosity"}\/) is much less than block
porosity $m$: fluid is contained in pores but moves through
cracks which are much wider than the pores but occupy much less
space.  Therefore
$$
q=\a m\big( \p_t h-  \frac{\rho gk}{\mu m_1} (\p_x
h)^2\big) \ . \eqno[5]
$$
Substitution of [5] to the water balance equation [1]
gives the equation
$$
m(1-\a)\p_t h=\frac{\rho g k}{\mu} (h\p^2_{xx}  h +\big(1-\a
\frac{m}{m_1}\big)(\p_xh)^2)  \ ,
$$
which can be reduced to the form
$$
\p_t h=\frac{\k}{m(1-\a)} (h\p^2_{xx} h-(c-1)
(\p_xh)^2)  \ ,  \eqno[6]
$$
where
$$
c=\a \ \frac{m}{m_1} \ . \eqno[7]
$$
Here there is a specially important point to be mentioned:
If $m_1=m$ (non-fissurized purely porous medium), then $c=\a$
and obviously $c$ is always less than one because the absorption
cannot exceed the available amount of fluid.
However, if the fluid is fissurized, $m_1$ can be substantially
less than $m$, and $c=\a m/m_1$ can be substantially larger than
one. Replacing $x$ by $x/\sqrt{\k /m(1-\a)}$ \ and leaving
previous notation $x$ for transformed space coordinate,
we reduce the basic equation to the canonic form
$$
\p_t h=h\p^2_{xx} h-(c-1)(\p_xh)^2 \ , \eqno[8]
$$
which will be investigated further.

Note that the proposed model has some common features with the
Mirzadzhan-Zadeh model (4) of filtration of the
gas-condensate mixture, but is not identical to this model which
leads to a different basic equation.

\section{Self-similar solutions}

We look for self-similar solutions with shrinking support and with
finite time to collapse (total annihilation): For sufficiently
large $c$ (in fact, see later, for $c>3/2$)\medskip
$$
h= A(t_0-t)^\l F\left(
\frac{x-x_0}{B(t_0-t)^\mu}\right) \ . \eqno[9]
$$
Here $x_0$ is the point where the solution is collapsing at
$t=t_0$.  We assume  the structure of the groundwater ``dome" is
symmetric, so that $F(\xi)$, $\xi=(x-x_0)/ B(t_0-t)^\mu$
is an even function. Here $A,B$ and $\l$ are constants,
but as we will see later, of a different nature. We will
determine the function $F(\xi)$ in the interval
$0\leq\xi\leq 1$, so that $F(\xi)\equiv 0$ at $\xi\geq 1$.
The quantity
$$
x_f=B(t_0-t)^\mu \eqno[10]
$$
is  the contracting half-width of the groundwater dome.
As is well known (5) for the degenerate parabolic
differential equations of the type under consideration, the
support of the solution remains compact if it is compact
initially. Furthermore, the function $F$ can be normalized
arbitrarily, so we can assume $A=B^2\mu$. Substitute
[9] into the basic equation [8]. Bearing in mind that the
coefficients of the resulting equation for $F(\xi)$ cannot
contain the time $t$ explicitly, we obtain
$\l=2\mu -1$, and the equation for $F(\xi)$ assumes the form:
$$
F \ \frac{d^2F}{d\xi^2}-(c-1)
\left(\frac{dF}{d\xi}\right)^2-\xi \ \frac{dF}{d\xi} +
\frac{2\mu-1}{\mu} \ F=0 \ . \eqno[11]
$$

We turn  now to the boundary conditions.
The first one is the condition of symmetry
$$
F'(0)=0 \ . \eqno[12]
$$

The second boundary condition follows from the continuity of the
groundwater level at the free boundary $x=x_f$:
$$
F(1)=0 \ . \eqno[13]
$$
The last boundary condition follows from the continuity of the
groundwater flux at the free boundary $x=x_f$. The solution close
to free boundary can be considered as quasi-steady:
$h=h(\zeta)$, \ $\zeta =x-x_f$. We obtain from [8]:
$$
-V \ \frac{dh}{d\zeta} =h \ \frac{d^2h}{d\zeta^2} -(c-1)
\left(\frac{dh}{d\zeta}\right)^2 \eqno[14]
$$
where \ $V=dx_f/dt$.

For the solutions of equation [11] having the quantity
$d(F^2)/d\xi$ equal to zero at $\xi=1-0$, which is needed to have
continuous flux at the free boundary, the quantity
$hd^2h/d\zeta^2$ tends to zero at $x=x_f-0$, so that at $x=x_f$
$$
V =(c-1) \frac{dh}{d\zeta} \ . \eqno[15]
$$

Bearing in mind that for the self-similar solution [9],
$V=dx_f/dt=-B\mu(t_0-t)^{\mu -1}$, and
$(dh/d\zeta)_{\zeta=0}=(\p_x h)_{x=x_f}$, we obtain the
relation
$$
B\mu(t_0-t)^{\mu-1}=-B\mu(t_0-t)^{\mu-1}(c-1)F\,'(1)
$$
from which the third boundary condition follows, to be satisfied
by the function $F$
$$
F\,'(1)=- \ \frac{1}{c-1} \ . \eqno[16]
$$
So, a nonlinear eigenvalue problem is obtained:
We have to find for the {\it second order} equation [11] in the
interval $[0,1]$ the solution satisfying {\it three}
boundary conditions [12],[13],[16], and the eigenvalue $\mu$.

We consider the special case when the function $F(\xi)$ has a
maximum at $\xi =0$. This allows one  to search
the  solution in the form of an expansion
$$
F(\xi)=a(1-\xi^2)+\sum^{\i}_{n=2} a_n(1-\xi^2)^n \ ,
$$
so that the terms of the sum (except the first one) do not
contribute to all three boundary conditions of the eigenvalue
problem. The result is unexpectedly simple:
$$
a= \frac{1}{2(c-1)} \ , \qquad
\mu =\frac{c-1}{2c-3} \ , \qquad
a_n=0 \ \ (n\geq 2) \eqno[17]
$$
so that, if $c> 3/2$, the solution to the nonlinear eigenvalue
problem is obtained in the form
$$
F= \frac{1}{2(c-1)} (1-\xi^2) \ , \qquad
\mu=\frac{c-1}{2c-3}  \eqno[18]
$$
and the self-similar solution under consideration is represented
by the relation
$$
h=\frac{1}{2(2c-3)}B^2(t_0-t)^{\frac{1}{2c-3}}
\left[ 1-
\frac{(x-x_0)^2}{B^2(t_0-t)^{\frac{2(c-1)}{2c-3}}}\right] \ .
\eqno[19]
$$

\section{Investigation of the self-similar solutions}

In spite of its very simple form, the solution [19] is a typical
self-similar solution of the second kind (see (6)).
The exponent $\mu=(c-1)/(2c-3)$ cannot be obtained using some
conservation laws, but only by solving a nonlinear eigenvalue
problem, and the constants $B$ and $t_0$, as well as $x_0$, are
obtained by matching the self-similar solution with the solution
to the Cauchy problem at the non-self-similar stage.
To a certain extent this problem is similar to the problem of the
evolution of a turbulent burst (see (6), section 10.2.4).
The solution [19] has essentially different behavior in various
intervals of the values of the absorption coefficient $c$:
$$
0 < c < 1; \qquad
1 < c < \textstyle{\frac 32} \ ;\qquad
\textstyle{\frac 32} < c \ . \eqno[20]
$$
The form [19] is appropriate for the last interval
$3/2 < c$ where the collapse time $t_0$ is finite. It is
instructive to investigate the limiting behavior of the solution
[19] at $c\to 3/2$ from above. Putting $c=3/2+\e$ ($\e >0$ is a
small parameter), we obtain $\mu =1/4\e+\frac 12$, so that
$$
(t_0-t)^\mu = t_0^{\frac{1}{4\e}+\frac 12}
\left(1-\frac{t}{t_0}\right)^{\frac{1}{4\e}+\frac 12} \ ,\qquad
(t_0-t)^{2\mu-1} = t_0^{\frac{1}{2\e}}
\left(1-\frac{t}{t_0}\right)^{\frac{1}{2\e}} \ ,
$$
and
$$
x_f(t)= Bt_0
^{\frac{1}{4\e}+\frac 12}
\left(1-\frac{t}{t_0}\right)^{\frac{1}{4\e}+\frac 12}  \ , \qquad
h(x_0,t) = \frac{B^2}{4\e} t_0^{\frac{1}{2\e}+1}
\frac{1}{t_0} \ . \eqno[21]
$$

\medskip\ni
Therefore if at $\e\to 0$ the quantity $4\e t_0$ tends to a
certain constant $\Theta$, and the quantity
$B^2 \,t_0^{\frac{1}{2\e}+1}$ to another constant which we denote
by $C^2\Theta$, the solution [19] tends to a finite limit:
$$
h=C^2 \ e^{-2t/\Theta} \left[
1-\frac{(x-x_0)^2}{C^2\Theta \ e^{-2t/\Theta}}\right] \ ,
\qquad
x_f = C\sqrt{\Theta} \ e^{-t/\Theta} \ . \eqno[22]
$$

\medskip
In the interval $1< c < 3/2$ the exponent $\mu$ becomes negative,
and it is convenient to replace $\mu$ by $-\mu$, and $t_0$ by
$-t_0$. Solution [19] may be represented in a different form
\medskip
$$
h=\frac{1}{2(3-2c)} B^2 (t_0+t)^{- \ \frac{1}{3-2c}}
  \left[
1-\frac{(x-x_0)^2}{B^2(t_0+t)^{- \ \frac{2(c-1)}{3-2c}}}\right] \ ,
\eqno[23]
$$
\medskip
so that $h(x_0,t)=h_{\max}(t)$ and $x_f$ decay with time
according to the power laws \medskip
$$
h_{\max}(t) =\frac{1}{2(3-2c)} B^2 (t_0+t)^{- \
\frac{1}{3-2c}} \ , \qquad
x_f(t)=B(t_0+t)^{- \ \frac{c-1}{3-2c}} \ . \eqno[24]
$$ \medskip
The time of collapse is infinite and $t_0$ becomes simply an
additive constant. In the limit $c\to 3/2$ from below,
$c=3/2-\e$, $\e>0$, we obtain $\mu=1/4\e-1/2$, and \medskip
$$
h_{\max}(t) =\frac{1}{4\e}B^2 t_0^{-\frac{1}{2\e}+1}
\left(1+\frac{t}{t_0}\right)^{-\frac{1}{2\e}+1}\dt
\frac{1}{t_0} \ ,  \qquad
x_f(t) = Bt_0^{-\frac{1}{4\e}+\frac 12}
  \left(1+\frac{t}{t_0}\right)^{-\frac{1}{4\e}+\frac 12} \ .
$$
Assuming again that at $\e\to 0$ the quantity  $4\e t_0$ tends to
a constant $\Theta$, and $Bt_0^{-\frac{1}{4\e}+\frac 12}$
tends to another constant $C\sqrt{\Theta}$, we obtain the same
limiting formula [22].

In the interval $0 < c < 1$ (weak absorption) the compact support
extends, not contracts, although slower than in the case of
``porous medium equation" $c=0$. In this special case $c=0$
solution [19] is reduced to a known self-similar solution fo the
first kind ((7),(8); see also (9),(6)).
The degenerate special case $c=1$ was considered previously; the
papers by J.~R.~King (10) and P.~Rosenau (11)
should be mentioned specially. It is instructive to compare the
results obtained above with those obtained in the paper by
B.~Meerson et al (12).

\section{Nonuniqueness of solutions of the Cauchy problem}

We consider solutions of equation [18] with initial condition
$$
h(x,0)=h_0(x) \qquad {\mb{for}} \ x\in{\mathbb R} \ .
\eqno [25]
$$
where $h_0(x)$ is a continuous function which is positive in an
interval $(x_L(0),x_R(0))$ and which vanishes elsewhere.
Let $x_L(t)$ and $x_R(t)$ be two continuous functions for $t\geq
0$ such that $x_L(t)$ is nondecreasing,  $x_R(t)$ is
nonincreasing, and  $x_L(t)\leq x_R(t)$ for $t\geq 0$. It is
known that if $c\geq 1$ for any such pair $x_L(t)$ and $x_R(t)$
there exists a solution $h(x,t)$ of the Cauchy problem [8],[25]
such that $h(x,t)$ is positive if $x_L(t)< x < x_R(t)$, $t\geq 0$
and $h(x,t)$ vanishes elsewhere. For the proof, the definition of
solution, and further references we refer to (13).

Of special interest is the  choice of steady interfaces:
$x_L(t)=x_L(0)$ and $x_R(t)=x_R(0)$ for all $t\geq 0$.
The corresponding solution is larger than any other solution, and
in (14) a numerical scheme has been introduced which leads to this
unique solution. From the modeling point of view it is interesting
to observe that this solution can be obtained from the following
limiting procedure: replace $h_0(x)$ by $h_0(x)+\e$ $(\e>0)$,
solve problem [8],[25], and let $\e \to 0$.

Angenent (15) has constructed a solution of [8],[25] if $h_0(x)$
has nonzero slope at $x_L(0)$ and $x_R(0)$ (for technical reasons
the construction is local in time), which is unique in the class
of solutions which can be expanded in a Taylor series of
suffcieintly high degree near the interfaces:
\[
h(x,t)=\sum^N_{k=0} c_k(t)(x-x_L(t))^k +o((x-x_L(t))^N) \quad
{\mb{as}} \ x\to x_L(t) \ ;
\]
a similar expression holds at the right interface
$x=x_R(t)$. Here the uniqueness not only refers to $h$, but also
to the interfaces. We observe that the self-similar solution [19]
belongs to this class of solutions.

In section 5 we shall construct a numerical scheme which yields
solutions converging to the self-similar solution for large times,
and it is natural to ask if these solutions belong to the class
introduced by Angenent. In particular this scheme yields solutions
which are different from the ones obtained by the scheme in (14).
We conjecture that the solutions which we construct in the
following section are physically relevant, but undoubtedly future
research is needed to provide definite answers to the nonuniquness
question.

\section{Numerical experiment}

The goal of the numerical experiment was to indicate that the
self-similar  solution obtained above attracts the solutions to
non-self-similar Cauchy problems having the initial condition of
compact support, generally speaking a non-symmetric one:
$$
h(x,0)= h_0(x) \ , \qquad
x_L(0)\leq x \leq x_R(0) \ ,
$$
and $h(x,0)\equiv 0$ outside the interval $x_L(0)\leq x\leq
x_R(0)$. The basic equation [8] can be transformed to a form
convenient for numerical calculations
$$
\p_t h=\frac{{\dot x}_0(t)-{\dot x_f}(t)\xi}{x_f(t)} \
\p_\xi h+\frac{1}{[x_f(t)]^2} \
[h\p^2_{\xi\xi}h-(c-1)(\p_\xi h)^2] \ ,
$$
where \ ${\dot x}_0=dx_0/dt$, \ ${\dot x_f}=dx_f/dt$, and
$$
\xi =\frac{x-x_0(t)}{x_f(t)} \ , \qquad
x_f(t)=\frac{x_R(t)-x_L(t)}{2} \ , \qquad
x_0(t)=\frac{x_R(t)+x_L(t)}{2} \ ,
$$
so that the interval of new space variable $\xi$ becomes fixed:
$-1\leq\xi\leq 1$, whereas the solution $h(x,t)$ is different from
zero in the time dependent interval
$x_L(t)\leq x\leq x_R(t)$. Again, assuming naturally the
quasi-steadiness of the level distribution in the vicinities of
free boundaries, we obtain the conditions
$$
{\dot x}_R(t)=(c-1)
\frac{\p_{\xi} h(1,t)}{x_f(t)} \  , \qquad
{\dot x}_L(t)=(c-1)
\frac{\p_{\xi}h(-1,t)}{x_f(t)} \eqno[26]
$$
at $t>0$, and the basic equation takes the form:
$$
\p_t h=
\frac{1}{x^2_f(t)} \left[(c\!-\!1)\p_\xi h
\frac{(\xi +1)\p_\xi h(1,t)\!-\!(\xi\!-\!1)\p_\xi h(-1,t)}{2}
+ h \p_{\xi\xi}h-(c\!-\!1)(\p_\xi h)^2 \right] \eqno[27]
$$
with the initial condition
$$
h(\xi,0) =h_0(\xi) \ , \quad
|\xi|\leq 1 \ ; \qquad h(\xi,0)\equiv 0 \ , \quad
|\xi|\geq 1 \ . \eqno[28]
$$

Two numerical schemes were used in our computations performed by
finite-difference approximations: (i) a forward-in-time,
centered-in-space {\it explicit} approximation, and (ii) a
forward-in-time,
centered-in-space {\it implicit} approximation.
For the most part numerical calculations have been run with the
time step $\Delta t=10^{-5}$ for the explicit scheme and
$\Delta t=10^{-4}$ for the implicit one. The number $N$ of
subintervals of length $\Delta\xi$, $N=2/\Delta\xi$ was equal to
202 for both schemes. The results obtained by using these
numerical approximations coincided with good accuracy.
The absorption coefficient $c$ was always equal to 1.75.

The first initial condition was taken as a ``smoothed block":
a homogeneous water level distribution smoothly going to zero at
the edges. The results of the computation are presented in Figure
1 in the form of the distribution of the scaled level: level
$h(x,t)$ divided by maximum level at each time $h_{\max}(t)$. It
is seen that the curves corresponding to different times collapse
to the parabola, corresponding to the self-similar solution [18].
The time of collapse $t_0$ and the constant $B$ were determined in
the following way (Figure 2): According to the intermediate
asymptotics [19] at small $t_0-t$,
$$
\frac{x^2_f(t)}{h_{\max}(t)} =
\frac{t_0-t}{\mu F(0)} \ . \eqno[29]
$$
so that the quantity $x^2_f(t)/h_{\max}(t)$ should be a linear
function of time, and the intersection of its graph with the time
axis (Figure 2,a) gives the value of $t_0$. We obtain from [10] a
linear relation between $\ln x_f$ and $\ln(t_0-t)$, i.e.~in the
coordinates $-\ln x_f,-\ln(t_0-t)$ a straight line with the slope
$\mu$. It gives us the value $B$ (Figure 2,b) and an additional
possibility of checking the asymptotics. Naturally $t_0$ and $B$
depend on the initial condition.
For our case we found $t_0=0.345$, $B=3.73$ and $\mu=1.499$, which
agrees well with the analytic value $\mu =1.5$.

The next computation was performed for a nonsymmetric initial
condition:
$$
h(x,0)=\begin{cases}
-4x^2+4x \ , & 0 < x < \frac 12 \\
-\frac 49 x^2 + \frac 49 x + \frac 89 \ , &
\frac 12 < x < 2 \ . \end{cases} \
$$

Figures 3,a and 3,b demonstrate the behavior of
the numerical solution for different times. It is clearly seen
that the solution becomes symmetric and tends to the self-similar
asymptotics [10]. The values of $t_0=1.138$, $B=0.63$ and
$\mu=1.5027$ have been calculated as before; the calculated and
analytic values of $\mu$ agree with high precision.

For comparison we have taken the solution [19] for a certain $t$ as
an initial condition, and computed the solution to the partial
  differential equation further using the same algorithm. The
results are presented in Figure 4,a for
different times. Being plotted in scaled coordinates
(Figure 4,b) they collapse to a single curve,
giving us an additional check of the numerical procedure.

\section{Conclusions}

We presented a new derivation of the filtration-absorption
equation based on a model of groundwater flow with partial
absorption. It is shown that for a sufficiently large absorption
constant the time of collapse is finite.
  A family of self-similar solutions to this equation
is obtained. Numerical experiments indicate that these
self-similar  solutions obtained  are self-similar intermediate
asymptotics for the solutions to the Cauchy problems having the
initial conditions data with compact support, but due to the
nonuniqueness of the solution of the Cauchy problem future
research is needed to provide more definite conclusions.

\bs
We express our gratitude to Professor A.~J.~Chorin and Professor
R.~Dal Passo for their valuable comments. We thank Professor
S.~Abarbanel for his interest in our work. This work was supported
in part by the National Science Foundation under Grant DMS
97--32710, and in part by the Applied Mathematics subprogram of
the U.S.~Department of Energy under contract DE--AC03--76--SF00098.

\bs\bs
\begin{enumerate}
\item Polubarinova-Kochina, P.~Ya. (1962). {\it
Theory of Groundwater Movement}, Princeton University Press,
Princeton.
\item Bear, J. (1972). {\it Dynamics of Fluids in Porous Media},
Dover, New York.
\item Barenblatt, G. I., Entov, V. M., and Ryzhik, V. M. (1990).
{\it Flow of Fluids Through Natural Rocks}, Kluwer Academic,
Dordrecht, The Netherlands.
\item Magerramov, N.~Kh., and Mirzadzhan-Zadeh (1960).
{\it Journal of Applied Mathematics \& Mechanics (Prikl. Mat.
Mekh.)}, vol.~XXIV, no.~6.
\item Kalashnikov, A. S. (1987). {\it Russian Math. Surveys} {\bf
42}, 169--222.
\item Barenblatt, G. I. (1996). {\it Scaling, Self-similarity, and
Intermediate Asymptotics}, Cambridge University Press, Cambridge.
\item Zeldovich, Ya.~B., and Kompaneets, A. S. (1950).
On the theory of propagation of heat with thermal conductivity
depending on temperature. In Collection of Papers Dedicated to the
70th Birthday of A.~F.~Ioffe, pp.61--71 {\it Izd.~Akad.~Nauk
USSR}, Moscow.
\item Barenblatt, G. I. (1952).  {\it Prikl. Mat. Mekh.} {\bf 16},
no.~1, 67--78.
\item Zeldovich, Ya.~B., and Raizer, Yu. P. (1967).
{\it Physics of Shock Waves and High Temperature Hydrodynamic
Phenomena}, vol.~1, Academic Press, New York.
\item King, J. R. (1993).
{\it Journ. Engng. Math.} {\bf 27}, no.~2, 31--72.
\item Rosenau, P. (1995).
{\it Phys. Rev. Letters} {\bf 74}, no.~7, 1056--1059.
\item Meerson, B., Sasorov, P.V., and Sekimoto, K. (2000)
{\it Phys. Rev. Letters E} {\bf 61}, no.~2, 1403--1406.
\item Bertsch, M., Dal Passo, R., and Ughi, M. (1992).
{\it Annali di Matematica pura ed applicata} (IV), vol.~CLXI,
57--81.
\item Bertsch, M. and Dal Passo, R. (1990). {\it Quarterly Appl.
Math.} {\bf 48}, 133--152.
\item Angenent, S. (1980). {\it Math. Annalen} {\bf 280}, 465--482.

\end{enumerate}

\newpage
\begin{center}{\bf Figure Captions}\end{center}

\ni {\bf Figure 1.} The numerical solution to the Cauchy problem
for
$c=1.75$ with the initial condition of a ``smoothed block" type
for different times in the scaled coordinates. The solution
is collapsing to the self-similar asymptotics.

\bs\ni {\bf Figure 2.}\\(a) The determination of the parameter
$t_0$.

\ni (b)  The determination of the parameter $B$.

\bs\ni {\bf Figure 3.}\\(a) Evolution of the nonsymmetric initial
distribution to a symmetric self-similar\\asymptotics.

\ni (b) The same evolution presented in scaled coordinates.

\bs\ni{\bf Figure 4.} The numerical solution preserves the
self-similarity.

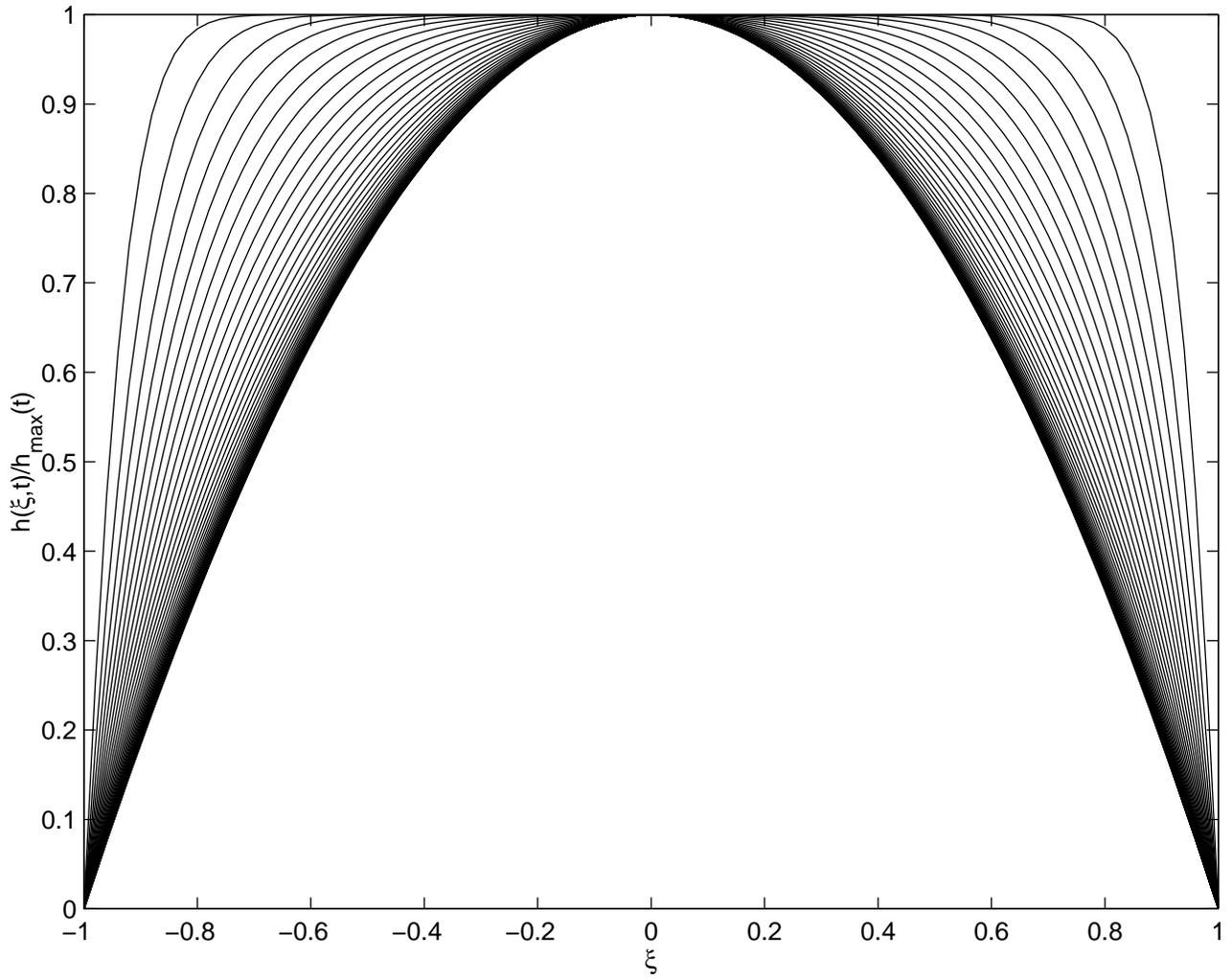
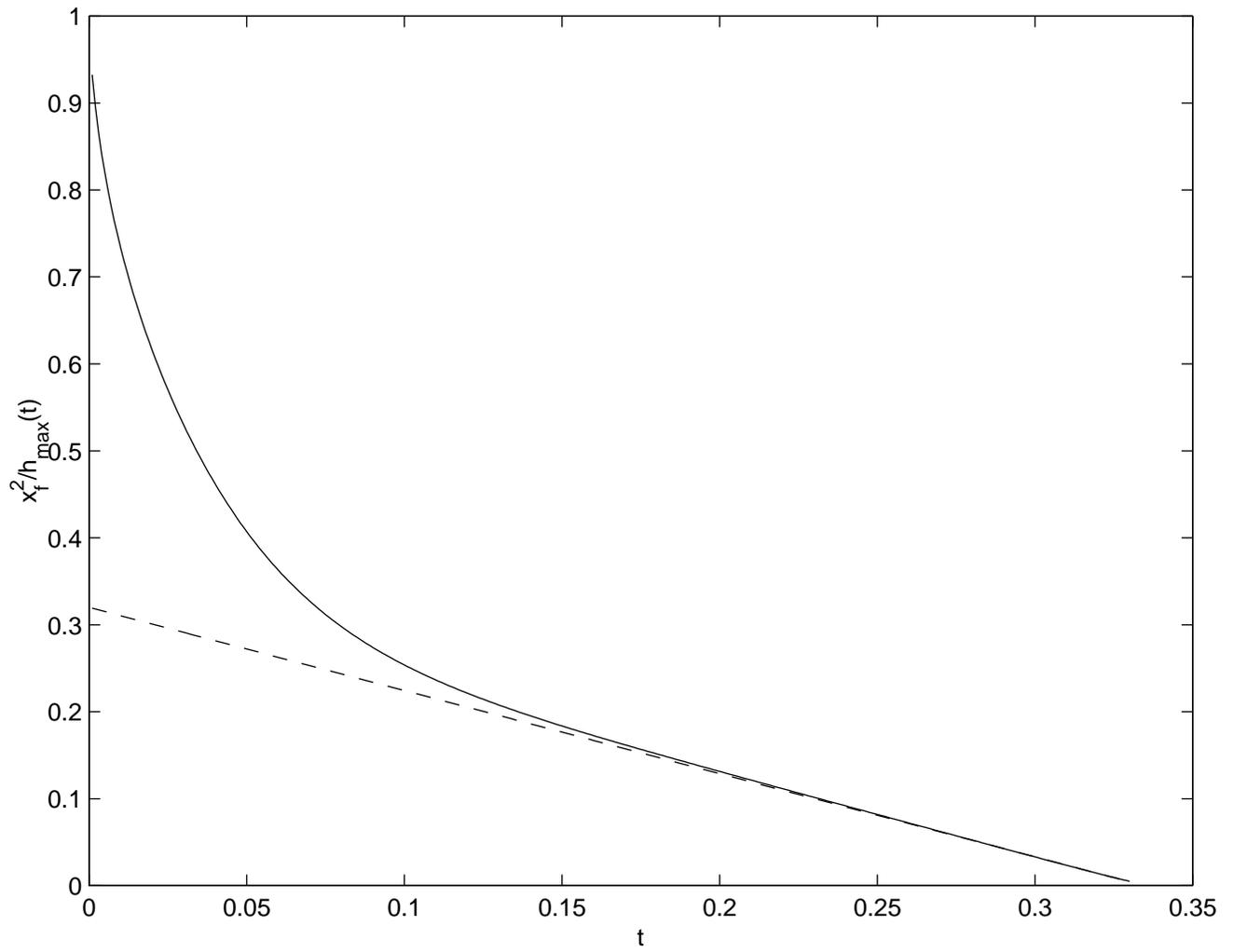
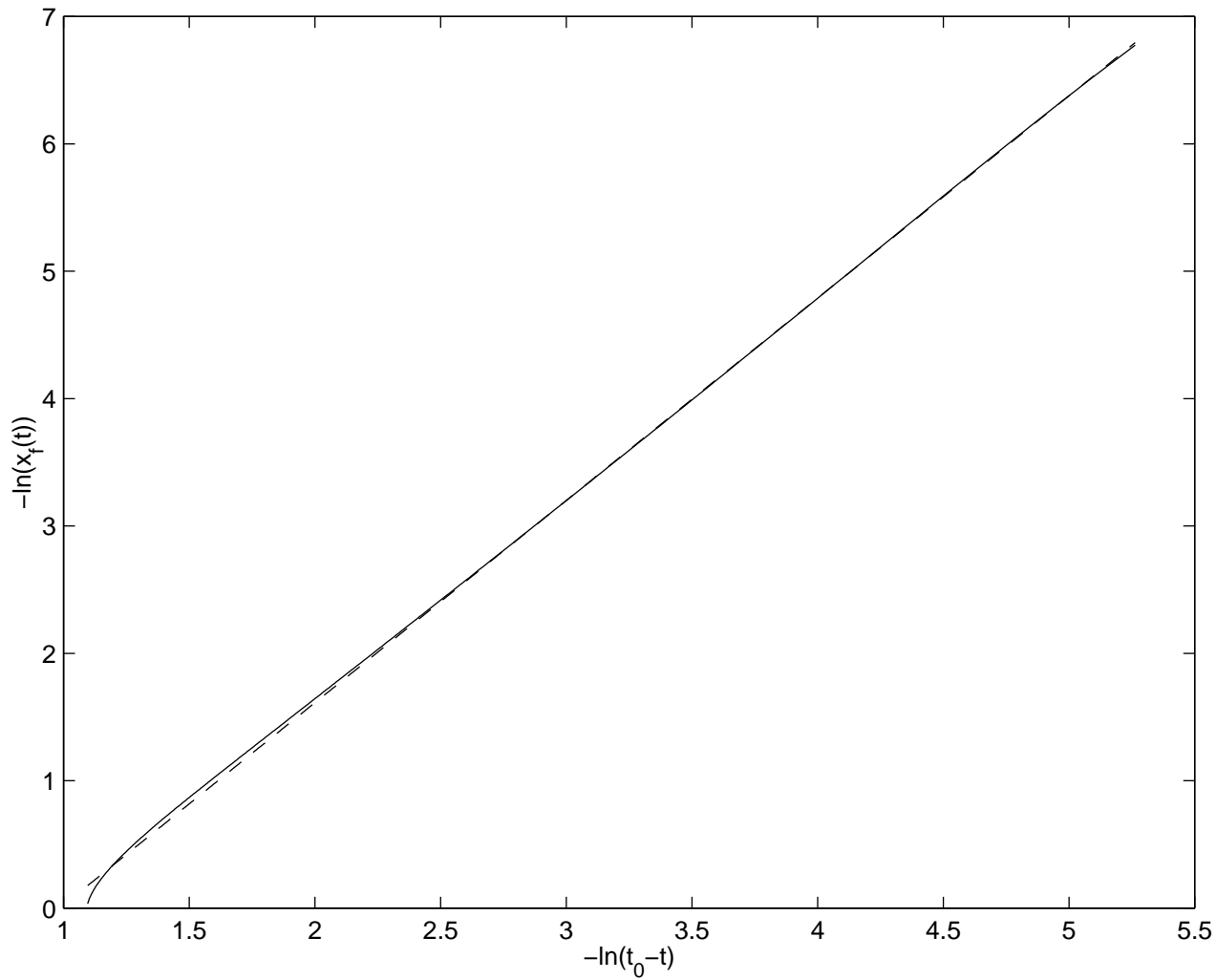
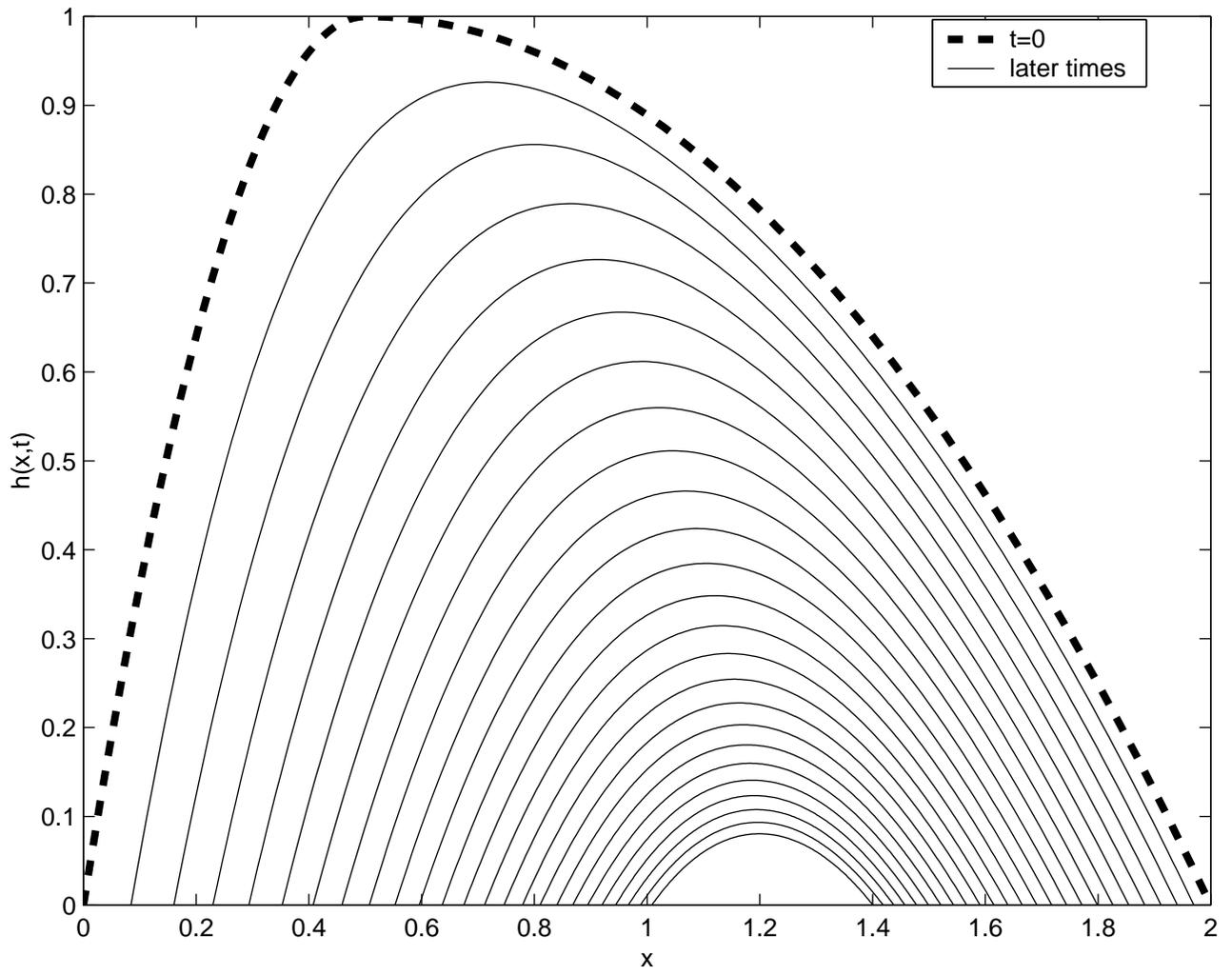
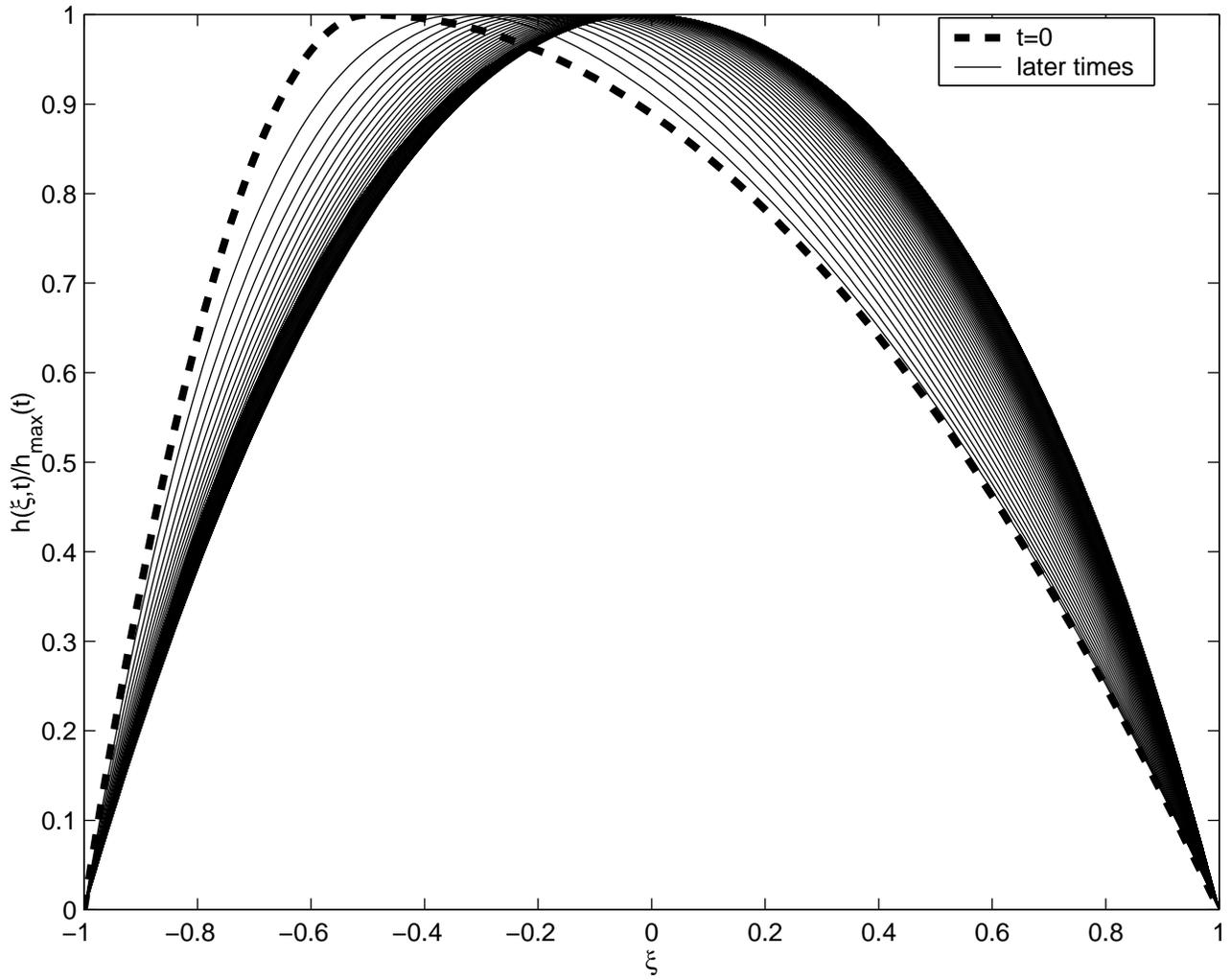
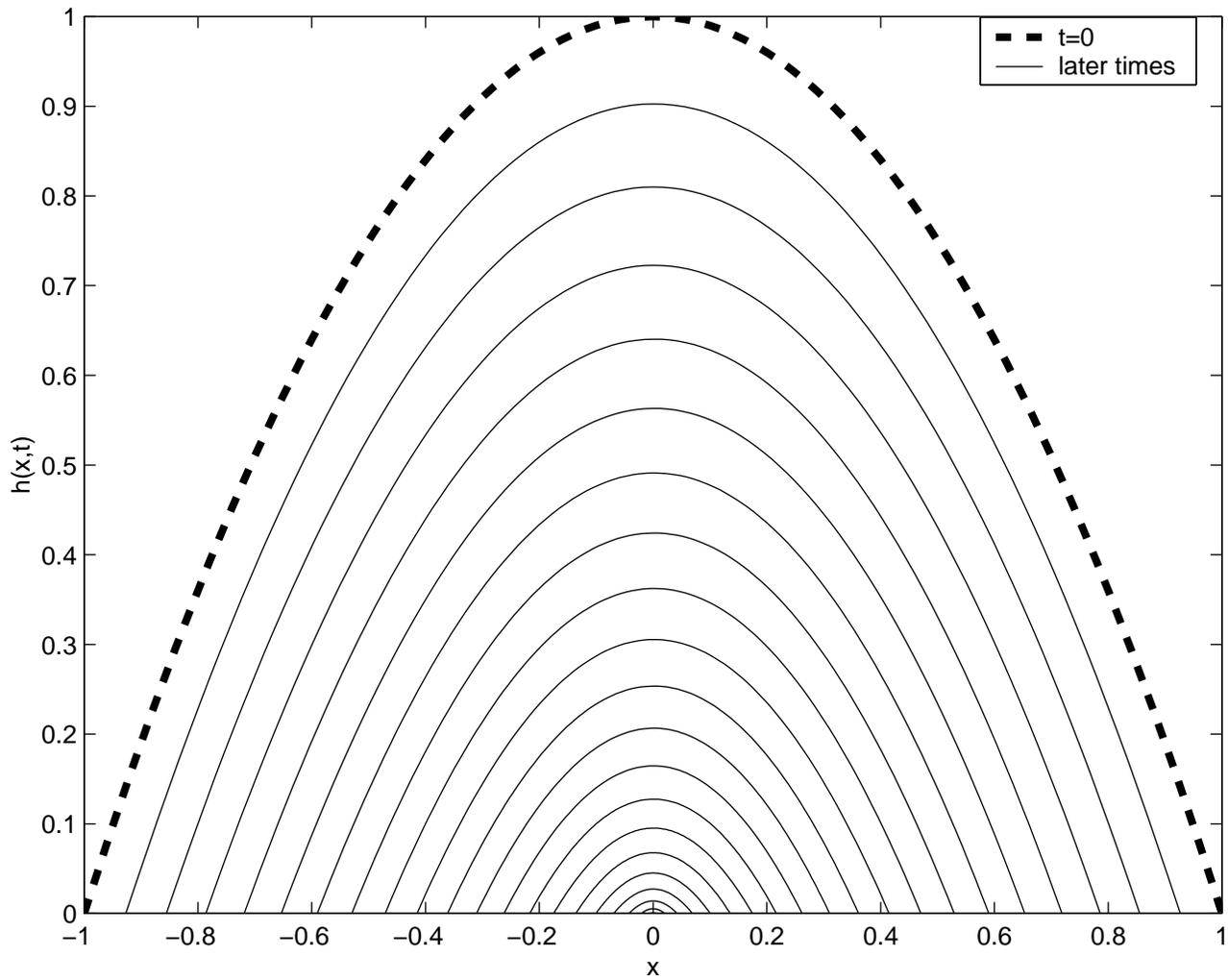
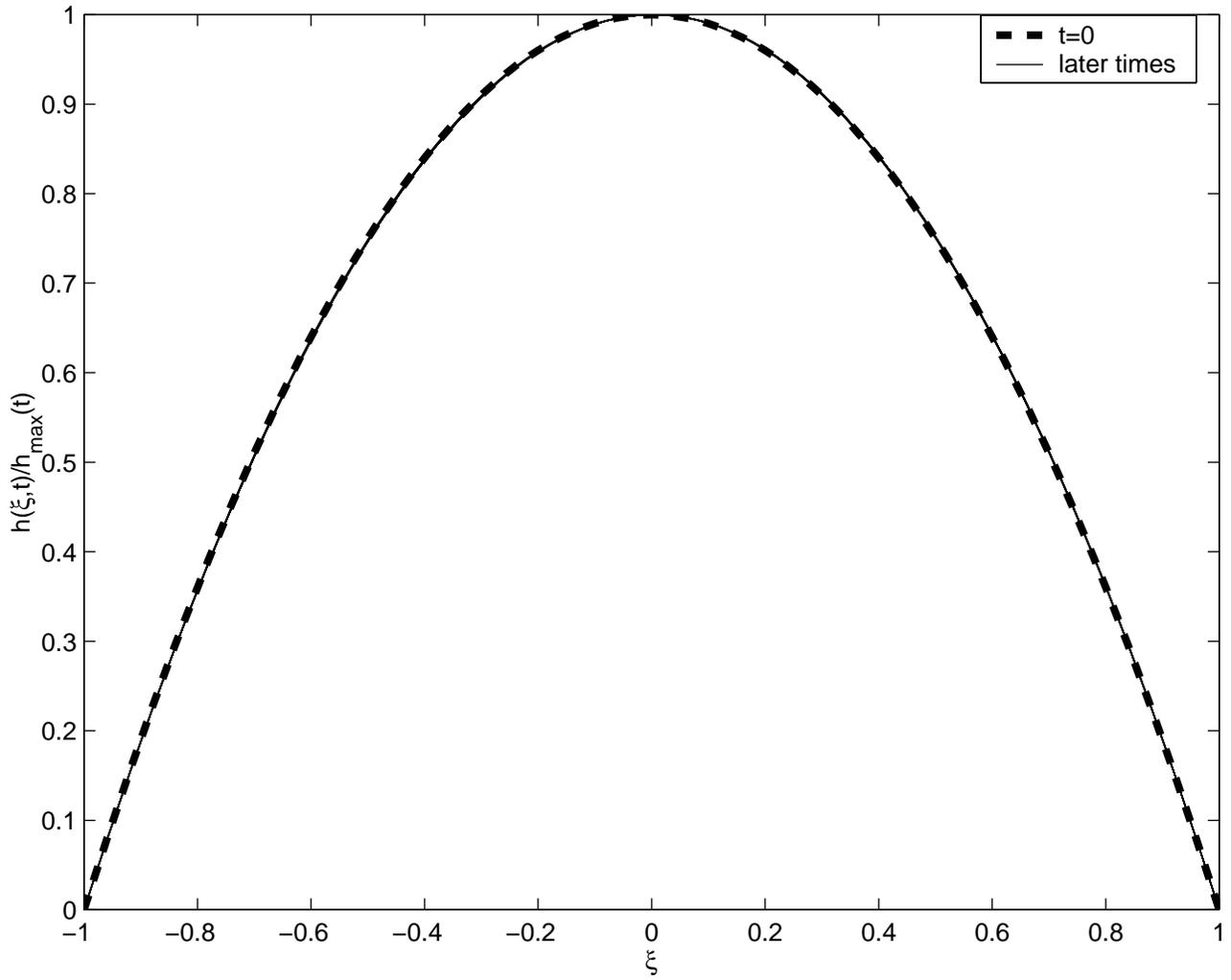
\end{document}